\documentclass[prb,twoside,twocolumn,floatfix,showpacs,citeautoscript,superscriptaddress]{revtex4-2}

\usepackage{amsmath}
\usepackage{amssymb}
\usepackage{booktabs}
\usepackage{glossaries}
\usepackage{graphicx}
\usepackage[version=3]{mhchem}
\usepackage{siunitx}
\usepackage{hyperref}

\newcommand \Title{
    Quantifying Dynamic Tilting in Halide Perovskites:\texorpdfstring{\\}{}
    Chemical Trends and Local Correlations
}

\hypersetup{
    pdfauthor={J. Wiktor, E. Fransson, D. Kubicki, and P. Erhart},
    pdftitle={\Title},
    pdffitwindow=false,
    pdfstartview={FitH},
    pdfnewwindow=true,
    colorlinks=true,
    linkcolor=black,
    citecolor=black,
    filecolor=black,
    urlcolor=black,
}

\graphicspath{{figures/}}

\setacronymstyle{long-short}
\newacronym{cx}{vdW-DF-cx}{van-der-Waals-density functional with consistent exchange}
\newacronym{dft}{DFT}{density functional theory}
\newacronym{gpu}{GPU}{graphics processing unit}
\newacronym{md}{MD}{molecular dynamics}
\newacronym{nep}{NEP}{neuroevolution potential}
\newacronym{paw}{PAW}{projector augmented-wave method}
\newacronym{scan}{SCAN}{strongly constrained and appropriately normed}
\newacronym{sm}{SM}{Supplemental Material}
\newacronym{sscha}{SSCHA}{stochastic self-consistent harmonic approximation}
\newacronym{wham}{WHAM}{weighted histogram analysis method}
\newacronym{xc}{XC}{exchange-correlation}

\DeclareSIUnit\angstrom{\text{Å}}
\renewcommand{\vec}[1]{\ensuremath\boldsymbol{#1}}

\newcommand{\addchalmers}{%
    Department of Physics, Chalmers University of Technology, SE-412 96 Gothenburg, Sweden
}
\newcommand{\addbirmingham}{%
    School of Chemistry, University of Birmingham, Birmingham, UK
}

\begin{document}
\title{\Title}
\author{Julia Wiktor}
\email{julia.wiktor@chalmers.se}
\affiliation{\addchalmers}
\author{Erik Fransson}
\affiliation{\addchalmers}
\author{Dominik Kubicki}
\affiliation{\addbirmingham}
\author{Paul Erhart}
\email{erhart@chalmers.se}
\affiliation{\addchalmers}

\begin{abstract}
Halide perovskites have emerged as one of the most interesting materials for optoelectronic applications due to their favorable properties, such as defect-tolerance and long charge carrier lifetimes, which are attributed to their dynamic softness.
However, this softness has led to apparent disagreements between the local instantaneous and global average structures of these materials.
In this work, we assess the local tilt angles of octahedra in the perovskite structure through large-scale molecular dynamics simulations using machine learned potentials based on density functional theory.
We compare structural properties given by different density functionals, namely PBE, PBEsol, SCAN, and vdW-DF-cx, and establish trends across a family of \ce{Cs$MX$3} with $M$=Sn or Pb and $X$=Cl, Br or I perovskites.
Notably, we demonstrate a strong short-range ordering that persists even in the cubic phase of halide perovskites.
This ordering is reminiscent of the tetragonal phase and bridges the disordered local structure and the global cubic arrangement.
Our results provide a deeper understanding of the structural properties of halide perovskites and their local distortions, which is crucial for further understanding their optoelectronic properties.
\end{abstract}

\maketitle

\section{Introduction}

Halide perovskites have gained significant attention as promising materials for various applications, including high-efficiency solar cells \cite{kojima2009organometal, kim2012lead, hodes2013perovskite}, lasers \cite{lei2021metal}, light-emitting diodes \cite{van2018recent}, and more \cite{kim2018halide}.
Their exceptional performance is attributed to various factors, such as their defect tolerance \cite{kang2017high, chen2019imperfections, kim2020defect} and long carrier lifetimes \cite{ponseca2014organometal, edri2014elucidating, hutter2017direct, crothers2017photon}.
These properties are linked to the high dielectric constants of these materials and effective screening of charges \cite{fabini2016dynamic, fu2021stereochemical}, which in turn can be traced back to their dynamic softness and ability to dynamically respond to the presence of excess charges.
Despite their importance, understanding and quantifying the underlying dynamics of halide perovskites remains challenging.

Several previous studies have highlighted the crucial role played by the dynamic local structure and octahedral tilting in halide perovskites \cite{adams2023classification}.
For example, some investigations have revealed that ignoring the local structure can lead to inaccurate interpretation of various experimental techniques, including X-ray absorption near edge structure spectroscopy \cite{cannelli2022atomic}, high energy resolution inelastic X-ray scattering \cite{beecher2016direct}, and pair distribution function analysis \cite{page2016short, bernasconi2018ubiquitous, morana2023cubic}.
Moreover, electronic structure calculations performed for simulations cells corresponding to the average perovskite structure have been shown to result in poor estimates of band gaps \cite{wiktor2017predictive, zhao2020polymorphous, wang2022accurate, morana2023cubic}, emphasizing the criticality of accounting for transient local distortions.
Moreover, octahedral tilting has been demonstrated to be vital in stabilizing the perovskite phase in \ce{FAPbI3} \cite{doherty2021stabilized}, further highlighting the significance of this property.
These findings underscore the importance of understanding the dynamic nature of halide perovskites and its implications for a wide range of applications.

Octahedral tilting, being one of the key properties of halide perovskites, has been assessed in several previous computational studies.
One of the simplest indicators of how much a materials will distort at finite temperatures can already be extracted from static calculations of potential energy surfaces or phonon dispersion curves at \SI{0}{\kelvin}, as has been done for instance in Refs.~\citenum{yang2017spontaneous, yang2020assessment}.
This way, one can also compare the performance of various functionals \cite{kaczkowski2021vibrational}, but it is not possible to predict the exact dynamic properties of a material from calculations done at \SI{0}{\kelvin} alone.
\Gls{md} simulations have also been used to study the dynamics of halide perovskites more directly, for instance to contrast different functionals \cite{kaiser2021first, FraWikErh2023}, to verify the effect of cation mixing \cite{ghosh2017good} or to understand the local disorder and its effect on experimental observations \cite{cannelli2022atomic, FraRosEri22}.
Various dynamic properties have been used in these studies to quantify the disorder, such as evolution of a single $M$--$X$--$M$ angle \cite{ghosh2017good}, distributions of $M$--$X$--$M$ tilt angles \cite{carignano2017critical, cannelli2022atomic} or distributions of halogen displacements \cite{kaiser2021first, carignano2017critical}.

In this work, we quantify the local octahedral tilting of halide perovskites through large-scale \gls{md} simulations and assess trends across functionals, temperatures, and material chemistries.
We use \gls{nep} models to accelerate \gls{md} simulations based on \gls{dft} calculations.
In the case of $\alpha$-\ce{CsPbI3}, we find that different functionals lead to similar trends in tilt angle distributions within the same phase, but drastically different phase transition temperatures.
Focusing further on the \gls{cx} functional, we assess the trends across the \ce{Cs$MX$} family of inorganic perovskites with varying halogen atoms ($X$=Cl, Br, and I) and metal cation ($M$=Sn, and Pb).
We finally create a bridge between the local environment of a single octahedron and the global perovskite structure through assessing the correlation between neighboring units.

\section{Methods}

\subsection{Machine-learned potential construction}

We constructed third-generation (NEP3) \gls{nep} models \cite{FanZenZha21, FanWanYin22} using a boot-strapping strategy, as described in detail in Ref.~\citenum{FraWikErh2023}.
In this process, we included the cubic ($Pm\bar{3}m$), two tetragonal ($I4/mcm$, and $P4/mbm$), and one orthorhombic ($Pnma$) structure, as well as the so-called delta-phase ($Pnma$).
The model construction was carried out using the \textsc{gpumd} package \cite{FanWanYin22} as well as the \textsc{calorine} package \cite{calorine} and \textsc{hiphive} packages \cite{EriFraErh19} for data preparation and analysis.
The performance of the final models is summarized in \autoref{tab:models} of the \gls{sm}.

\subsection{Density functional theory calculations}

To generate input data for model construction, we performed \gls{dft} calculations using the \gls{paw} method \cite{Blo94, KreJou99} as implemented in the Vienna ab-initio simulation package \cite{KreHaf93, KreFur96}.
Our calculations use a $\Gamma$-centered $\vec{k}$-point grid with density of \SI{0.18}{\per\angstrom} and a Gaussian smearing width of \SI{0.1}{\electronvolt}.
Four different exchange-correlation functionals were used, namely PBE \cite{PerBurErn}, PBSEsol \cite{PerRuz2008}, \gls{scan} \cite{SunRuzPer15}, and the \gls{cx} method \cite{DioRydSch04, BerHyl2014}.
The valence electron configurations used in the \gls{paw} datasets are provided in \autoref{tab:paw-setups} in the \gls{sm}.
We used a plane-wave energy cutoff to \SI{520}{\electronvolt} for all materials.

\subsection{\texorpdfstring{\Gls{md}}{MD} simulations}

\Gls{md} simulations were carried out using the \textsc{gpumd} package \cite{FanPerWan15, FanWanYin22}.
During the simulations the temperature was continuously increased from 20 to up to \SI{620}{\kelvin} over a period of \SI{100}{\nano\second}, yielding to a heating rate of \SI{6}{\kelvin\per\nano\second}.
For some compounds with low transition temperatures we terminated the simulations already at 420 or \SI{520}{\kelvin}.
The simulations were run in the $NpT$ ensemble, with temperature and pressure controlled using stochastic velocity \cite{BusDonPar07} and cell rescaling \cite{BerBus20}.
The time step is set to \SI{5}{\pico\second}, and the pressure to zero.
The simulation cell comprised $16\times 16\times12$ repetitions of the primitive unit cell of the orthorhombic perovskite structure, corresponding to \num{61440} atoms.

\subsection{Analysis of tilt angles}
To extract the tilt angle for each \ce{$MX$6} octahedron we applied polyhedral template matching using \textsc{ovito} \cite{Stukowski2010}.
The orientation of the octahedra is then represented by Euler angles as shown in \autoref{fig:functionals}.
This is done at each step of the heating runs, giving us access to the angle distribution as a practically continuous function of temperature.

\section{Results}

\subsection{Comparison of functionals}

\begin{figure*}
    \centering
    \includegraphics[width=\linewidth]{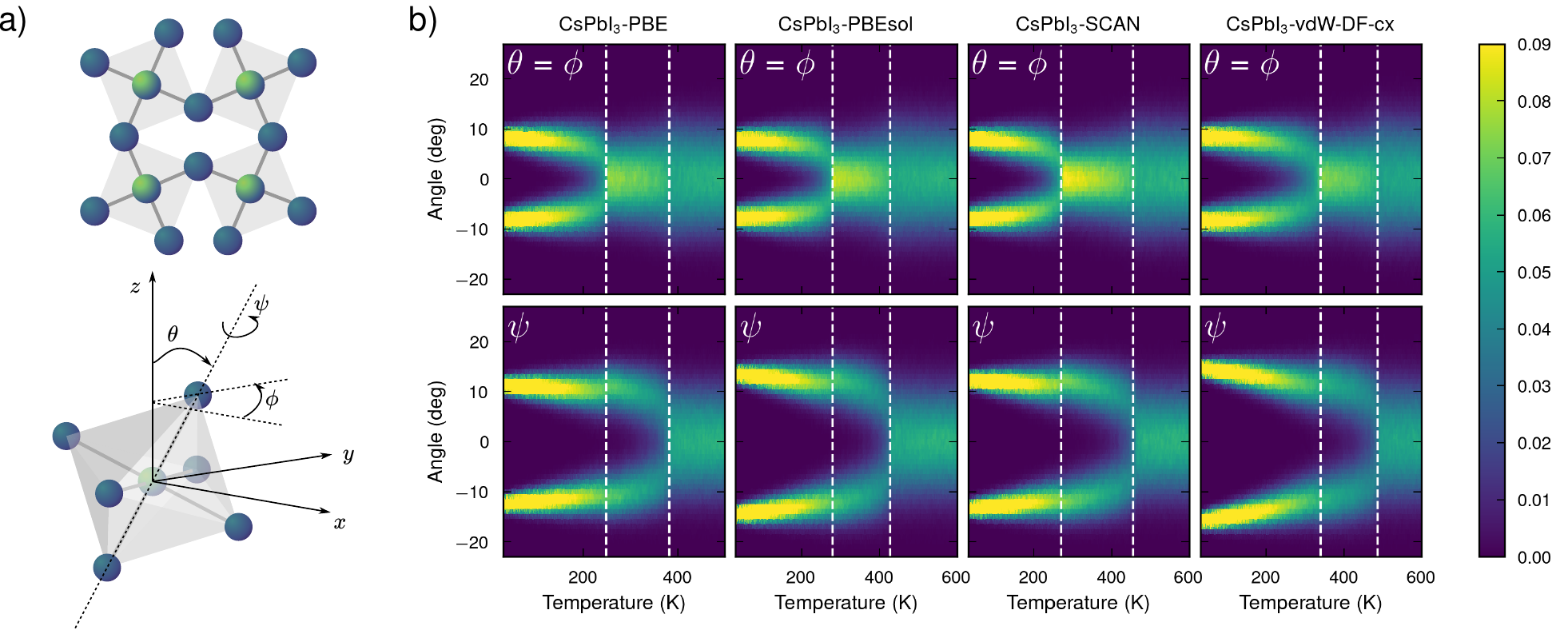}
    \caption{
        (a) Representation of the \ce{CsPbI3} perovskite structure with \ce{PbI6} octahedra indicating the definition of the three Euler angles $\theta$, $\phi$, and $\psi$ describing the orientation of the octahedron.
        (b) Maps of tilt angles as a function of temperature in PBE, PBEsol, SCAN, and \gls{cx} functionals.
        Dashed vertical lines indicate the orthorhombic-to-tetragonal and tetragonal-to-cubic phase transitions.
    }
    \label{fig:functionals}
\end{figure*}

We first compare the distributions of octahedral tilt angles computed for \ce{CsPbI3} according to \gls{nep} models representing four different functionals: the widely used PBE functional \cite{PerBurErn}, its variation adapted to modelling of solids, PBEsol \cite{PerRuz2008}, the \gls{scan} functional \cite{SunRuzPer15}, which has shown good performance for halide perovskites in comparison with calculations based on the random phase approximation \cite{bokdam2017assessing}, and the \gls{cx} method \cite{DioRydSch04, BerHyl2014}.
The distributions for all temperatures are shown in \autoref{fig:functionals} as maps of tilting angles.
The transition temperatures from the orthorhombic ($Pnma$) to the tetragonal ($P4/mbm$) phase at lower temperatures and from tetragonal to cubic ($Pm\bar{3}m$)) at higher temperatures as obtained in Ref.~\citenum{FraWikErh2023} are also marked.
Phase transition temperatures based on the models for all four functionals are given in \autoref{tab:phase-transitions-CsPbI3} and compared to experimental values.

\begin{table}
\caption{
    Phase transition temperatures in \si{\kelvin} found using different functionals for \ce{CsPbI3}.
    Values were extracted from the temperature dependence of the heat capacity \cite{FraWikErh2023}.
    Experimental values are given for comparison \cite{marronnier2018anharmonicity}.
}
\label{tab:phase-transitions-CsPbI3}
\centering
\begin{tabular}{l c c}
\toprule
& orthorhombic--tetragonal
& tetragonal--cubic \\
\midrule
PBE        & 249 & 382 \\
PBEsol     & 279 & 428 \\
\gls{scan} & 270 & 455 \\
\gls{cx}   & 340 & 487 \\
\midrule
experiment & 457 & 554 \\
\bottomrule
\end{tabular}
\end{table}

At low temperatures, the angle splitting are similar for all models (functionals).
The $\theta=\phi$ angle, characterizing the tilt in the $z$ direction in the orthorhombic structure has a magnitude of just below \SI{10}{\degree}.
The $\psi$ angle, related to the octahedral rotation in the $x-y$ plane is slightly larger than \SI{10}{\degree}.
The temperature evolution of the angles is also qualitatively similar, with the main difference being related to phase transition temperatures (see \autoref{tab:phase-transitions-CsPbI3}), with \gls{cx} giving the highest transition temperatures (the closest to experimental values) and PBE giving the lowest.
We also note that the $\psi$ angle in the orthorhombic phase has a stronger temperature dependence for the models based on the PBEsol and \gls{cx} functionals, than in the case of PBE and \gls{scan}.

\begin{figure*}
    \centering
    \includegraphics[width=0.98\linewidth]{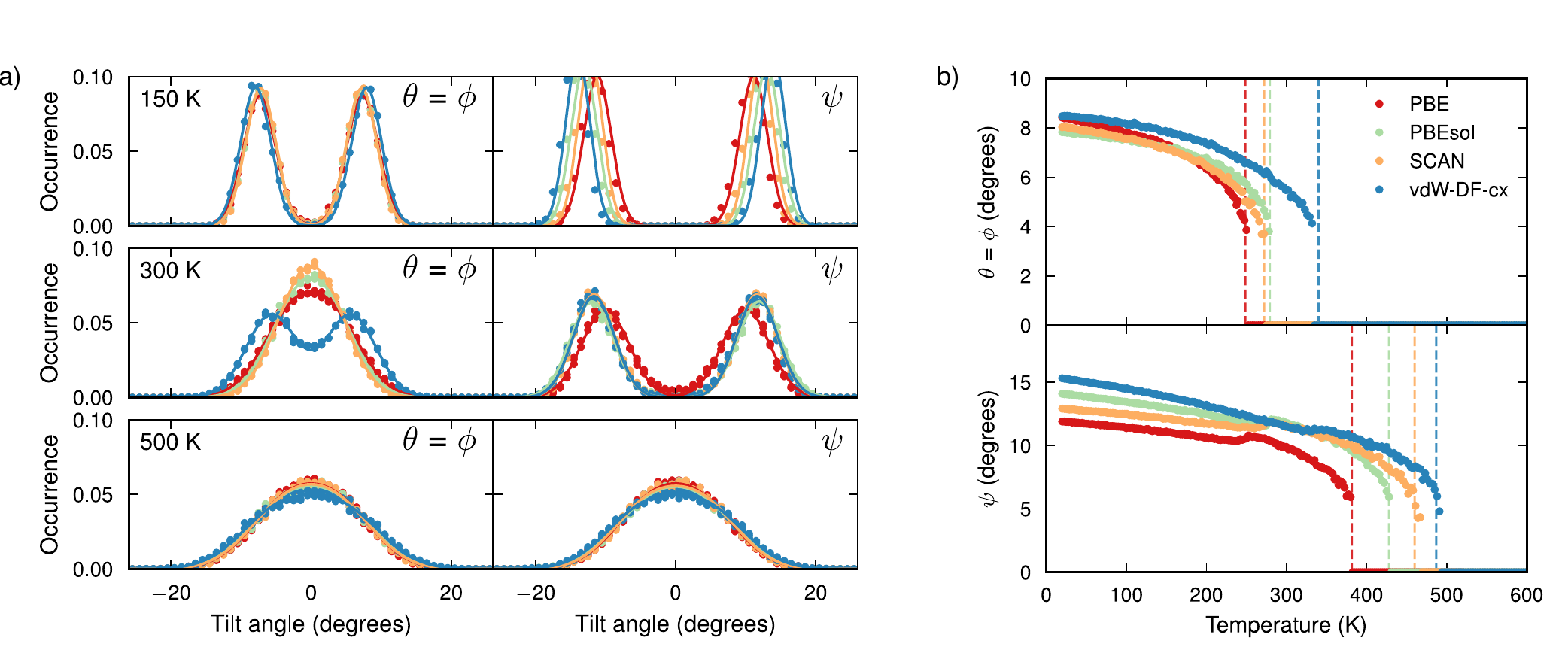}
    \caption{
        (a) Probability distribution of octahedral tilts in \ce{CsPbI3} as described by the three Euler angles $\theta$, $\phi$, and $\psi$.
        Data obtained with \gls{nep} models based on four different functionals (PBE, PBEsol, \gls{scan}, and \gls{cx}) are shown at three different temperatures.
        Solid lines are double Gaussian fits to the data.
        (b) Position of the maximum in the tilt angle distribution as extracted from double Gaussian fits.
        Vertical dashed lines indicate phase transitions.
    }
    \label{fig:funct2}
\end{figure*}

To enable a more straightforward comparison of angular distributions between the models representing the different functionals, we now analyze tilting at three selected temperatures, 150, 300, and \SI{500}{\kelvin} (\autoref{fig:funct2}).

At \SI{150}{\kelvin}, all models (functionals) correctly predict \ce{CsPbI3} to exhibit the orthorhombic structure.
While the $\theta=\phi$ distributions are almost the same for all functionals, the separation of the maxima in the $\psi$ distribution clearly increases in the order PBE$-$SCAN$-$PBEsol$-$\gls{cx}.

Next we focus on \SI{300}{\kelvin}, corresponding to the most commonly value used in \gls{md} simulations for halide perovskites.
At this temperature, \ce{CsPbI3} should exhibit the orthorhombic structure, according to experiments \cite{marronnier2018anharmonicity}.
Strikingly, among the considered functionals, only the \gls{cx}-based model predicts this phase to be stable at \SI{300}{\kelvin}, which is reflected in angle splitting for all three Euler angles.With the PBE, PBESol, and \gls{scan}-based models we observe $\theta=\phi$ distributions of similar broadening with one maximum only, indicating the tetragonal phase.
For the $\psi$ angle, all functionals yield a bimodal distribution.
The splitting of the maxima is very similar for all models (functionals) except for PBE, in which it is significantly smaller.
The differences between the distributions of $\theta=\phi$ achieved with different functionals highlight the importance of choosing the right computational setting for constant-temperature \gls{md} simulations.
We thus expect that room-temperature simulations of \ce{CsPbI3} in the orthorhombic phase using PBE, PBESol or \gls{scan} could lead to spurious distortions in the cell due to the drive towards the tetragonal phase.

At \SI{500}{\kelvin}, all functionals predict the same cubic phase, leading to similar angular distributions.
For all three Euler angles, we observe a broad Gaussian distribution centered around \SI{0}{\degree}.
We note that such a broad distribution of octahedral tilting angles is the reason for a large difference in the local and global symmetry in the halide perovskites \cite{cannelli2022atomic, wiktor2017predictive}.
While on average the structure at high temperatures exhibits a high-symmetry cubic structure, locally each octahedron is likely to be significantly tilted, with a standard deviation (see \autoref{fig:std} in \gls{sm}) from the average position of about \SI{7}{\degree} in each direction.

To further analyze the data, we fit the angle distributions with a symmetric double Gaussian function of the form
\begin{align*}
    p(x) = \frac{1}{2\sigma\sqrt{2\pi}} \left ( \text{e}^{-\frac{1}{2}(x-\mu)^2/\sigma^2} + \text{e}^{-\frac{1}{2}(x+\mu)^2/\sigma^2} \right ),
\end{align*}
where $\mu$ and $\sigma$ are mean and standard deviation respectively.
The fitting is done with a small regularization for the free parameters $\mu$ and $\sigma$ to specifically reduce noise when fitting the distributions from the cubic phase.

The positions of the maxima in the distributions ($\mu$) as a function of temperature are plotted in \autoref{fig:funct2}b.
The figure indicates that the splitting of the tilt angles exhibits a similar temperature dependence for all models (functionals).
We also note that the abrupt change in the positions of the maxima in the distributions agrees very well with phase transition temperatures (indicated by dashed lines in the figure) extracted from heat capacities \cite{FraWikErh2023}.
We also notice a clear correlation between the magnitude of the splitting in the angular distribution at low temperatures and the transition temperatures, with higher splittings corresponding to higher transition temperatures.

\subsection{Chemical trends}

\begin{figure*}
    \centering
    \includegraphics[width=0.98\linewidth]{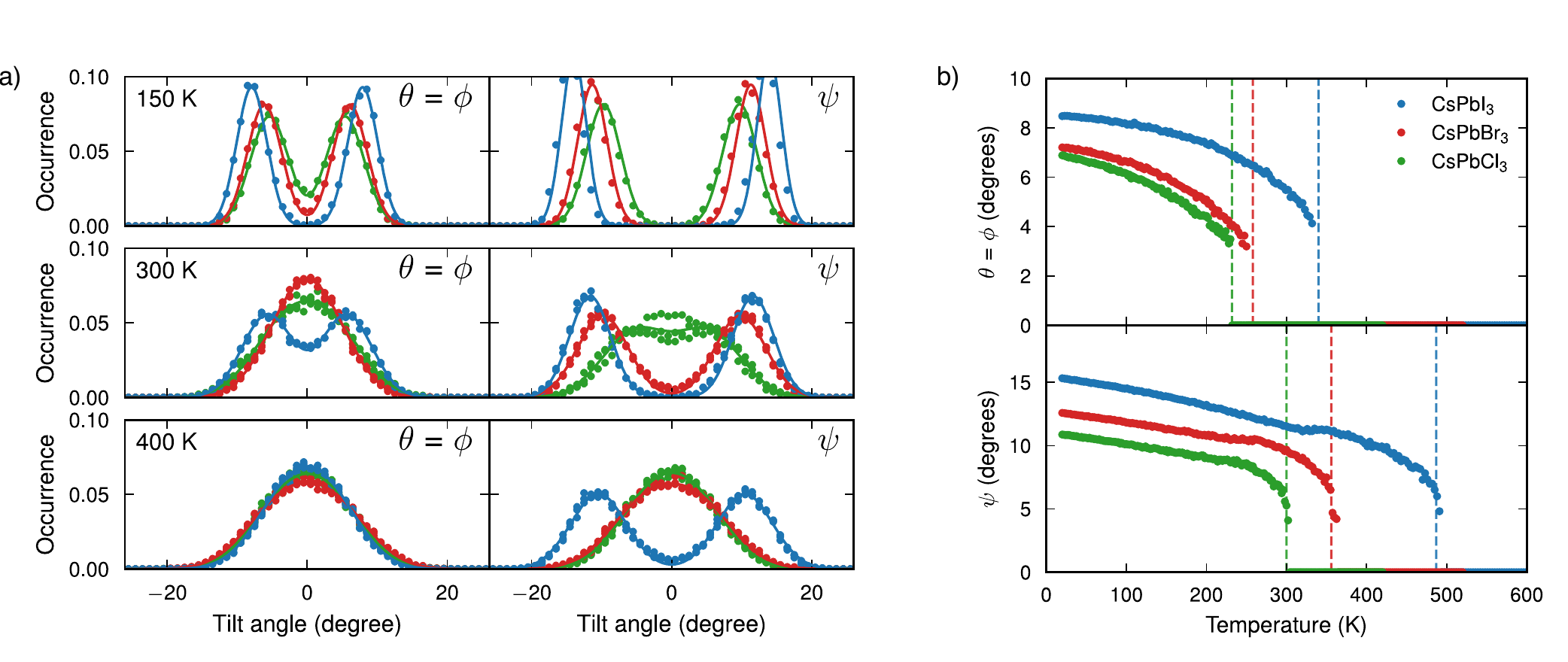}
    \caption{
        (a) Probability distribution of octahedral tilts at 150, 300, and \SI{400}{\kelvin} in \ce{CsPbCl3}, \ce{CsPbBr3}, and \ce{CsPbI3}, as described by the three Euler angles $\theta$, $\phi$, and $\psi$.
        (b) Position of the maximum in the tilt angle distribution as extracted from double Gaussian fits.
        Vertical dashed lines indicate phase transitions.
    }
    \label{fig:halides}
\end{figure*}

We now focus on the \gls{cx} functional, which exhibits the closest agreement with experiment with respect to transition temperatures and thermal expansion \cite{FraWikErh2023}, and compare different compounds.
We first analyze the effect of varying the halogen atom within the family of compounds \ce{CsPb$X$3} with $X=$Cl, Br, and I.
Tilting angle maps as well as transition temperatures can be found in the \gls{sm}, while in \autoref{fig:halides} we show the distributions at three temperatures (150, 300, and \SI{400}{\kelvin}) as well as the positions of maxima in the angular distributions.
At low temperatures, all three materials exhibit the orthorhombic $Pnma$ phase consistent with bimodal distributions of all three Euler angles.
The splitting between the maxima of both the $\theta=\phi$ and $\psi$ angles increases with the size of the halogen ion, following the sequence Cl--Br--I.
As can be seen in \autoref{fig:halides}b, this ordering is retained at all temperatures.

At \SI{300}{\kelvin}, \ce{CsPbCl3} undergoes a phase transition from the tetragonal to the cubic phase, which explains the mixed nature of the $\psi$ angle distribution.
\ce{CsPbBr3} is stable in the tetragonal phase, with a unimodal distribution of $\theta=\phi$ and a bimodal distribution of $\psi$.
\ce{CsPbI3} is still orthorhombic at \SI{300}{\kelvin} and exhibits bimodal distribution of all Euler angles.
We note that for \ce{CsPbCl3}, \SI{300}{\kelvin} is exactly the phase transition temperature according to the \gls{cx}-based model, which explains the mixed tetragonal/cubic distribution of the $\psi$ tilting angle.
At \SI{400}{\kelvin}, \ce{CsPbI3} is tetragonal, while both \ce{CsPbBr3} and \ce{CsPbCl3} are cubic.
This allows for a direct comparison between the later two.
While the plots in \autoref{fig:halides}a are rather similar, the distribution for \ce{CsPbCl3} is slightly more narrow, suggesting that the octahedral tilting is reduced for smaller halogen ions.

\begin{figure*}
    \centering
    \includegraphics[width=0.98\linewidth]{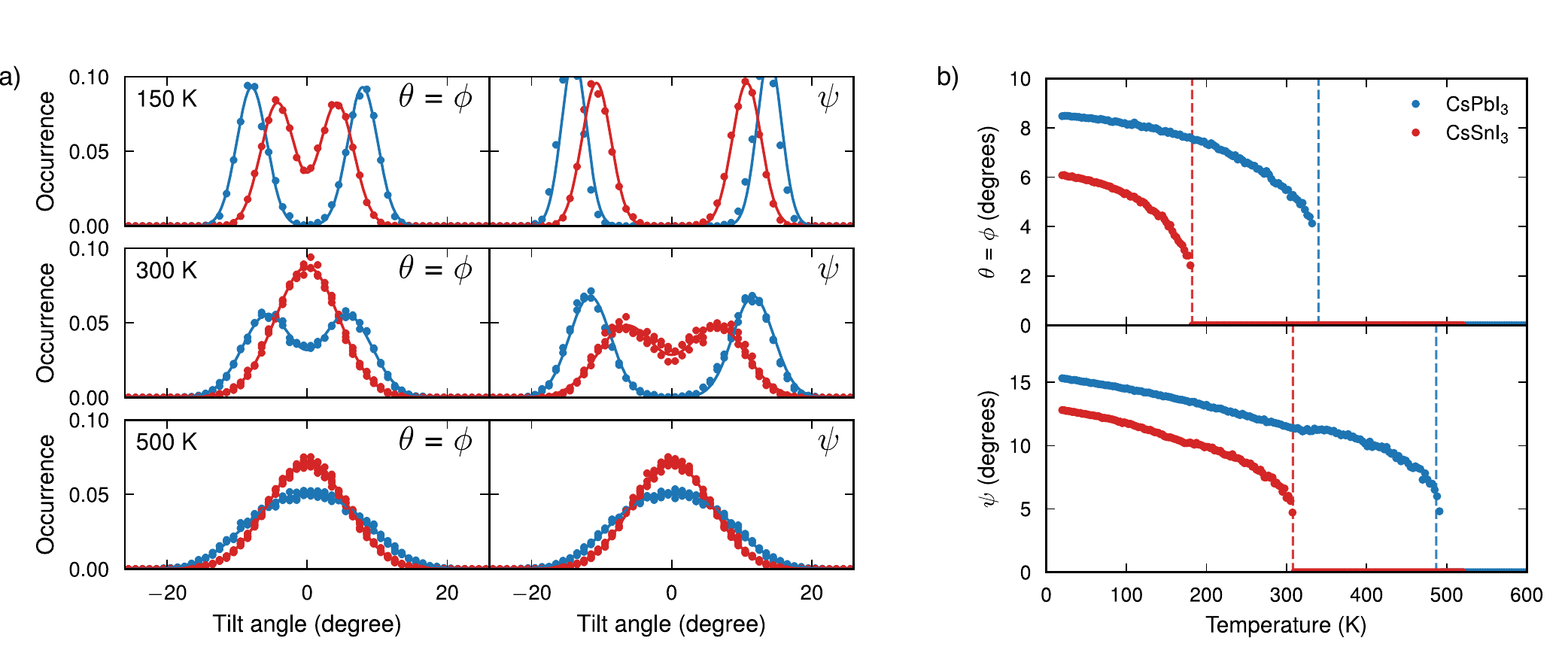}
    \caption{
        (a) Probability distribution of octahedral tilts 150, 300, and \SI{500}{\kelvin} in \ce{CsSnI3} and \ce{CsPbI3} compounds, as described by the three Euler angles $\theta$, $\phi$, and $\psi$.
        (b) Position of the maximum in the tilt angle distribution as extracted from double Gaussian fits.
        Vertical dashed lines indicate phase transitions.
    }
    \label{fig:metals}
\end{figure*}

We now evaluate the effect of the metal cation, by making a comparison between \ce{CsPbI3} and \ce{CsSnI3} (\autoref{fig:metals}).
At \SI{150}{\kelvin}, both \ce{CsPbI3} and \ce{CsSnI3} adopt the orthorhombic phase.
The tilt angle splitting is much more pronounced in \ce{CsPbI3} than in \ce{CsSnBr3}.
The smaller magnitude of tilt angles at low temperatures for \ce{CsSnI3} is correlated with much lower phase transition temperatures as compared to \ce{CsPbI3} (see \autoref{tab:transition-temperatures} in the \gls{sm}).

\subsection{Local tilt angle correlations}

\begin{figure*}
    \centering
    \includegraphics{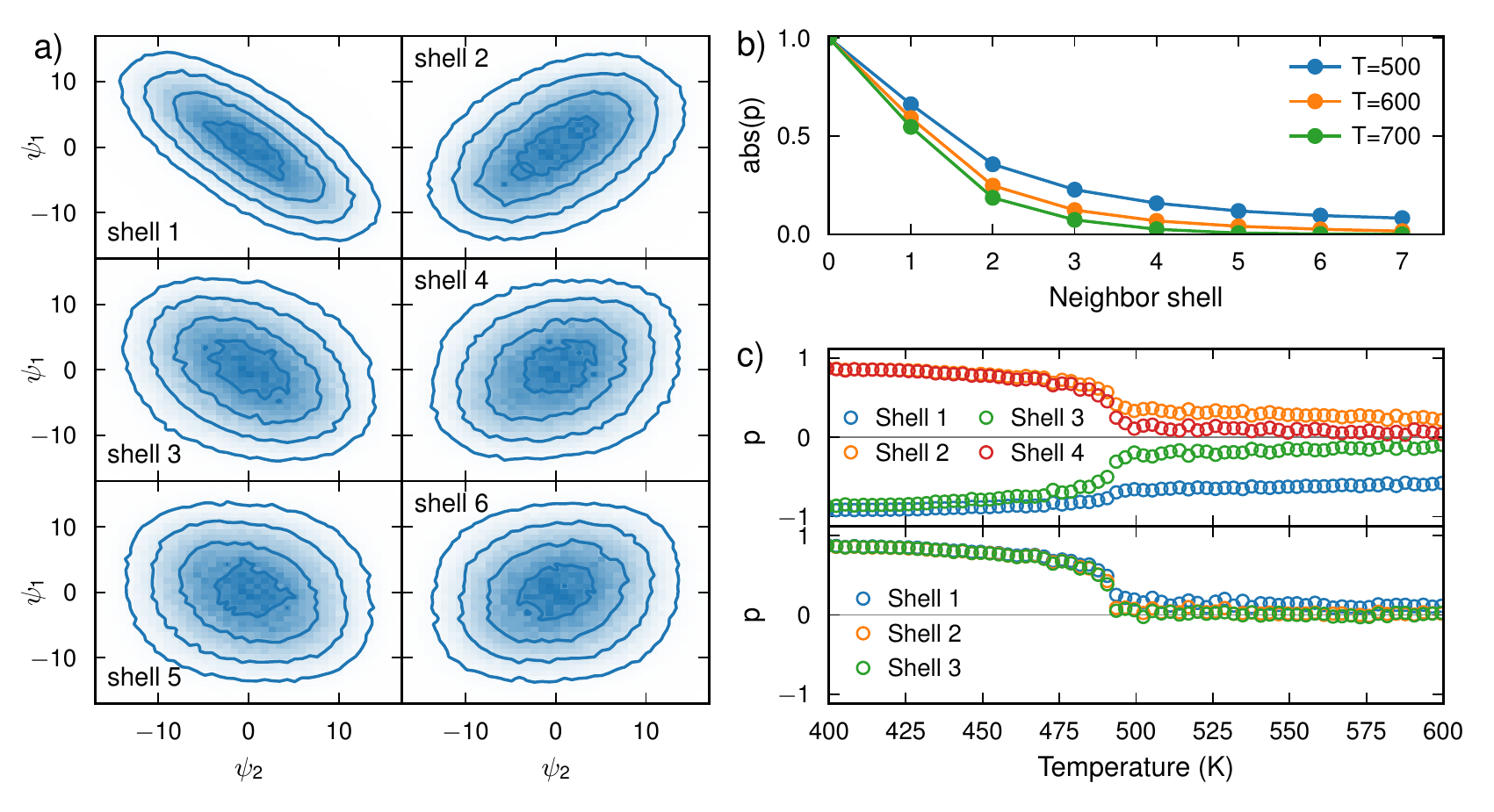}
    \caption{
        (a) Joint probability distribution $P(\psi_1, \psi_n)$ for tilt angles between octahedra in different neighbor shells along the [100] direction at \SI{500}{\kelvin} for \ce{CsPbI3}.
        (b) Correlation $|p|$ as a function of neighbor shell for 500, 600, and \SI{700}{\kelvin} along the [100] direction.
        (c) Correlation $p$ as a function of temperature along the [100] (top) and [001] (bottom) directions, where [001] is the rotation axis for $\psi$.
    }
    \label{fig:corr}
\end{figure*}

From analyzing the tilt angle distributions, it is clear that at high temperatures halide perovskites are cubic on a global scale corresponding to $\mu=0$.
However, locally  each octahedron can still be significantly tilted, with tilt angles reaching the values seen in the tetragonal and orthorhombic phases.
In the tetragonal and orthorhombic phases long-range ordering occurs with respect to the octahedral tilt angles.
In the cubic phase there is no long-range order, this does, however, not imply the absence of short-range order (correlation) between neighboring octahedra.
In the context of local regions in the cubic phase appearing tetragonal or orthorhombic \cite{cannelli2022atomic}, it is thus of interest to analyze and quantify the short-range order, i.e., assess the correlation between tilt angles of neighboring octahedra, in the cubic phase.
Strikingly, a recent study by Doherty \textit{et al.} found that phases that in X-ray diffraction appear to be cubic, locally feature regions that are tetragonal or orthorhombic \cite{doherty2021stabilized}.
Here, we investigate the joint distribution of the tilt angle of neighboring octahedra, $P(\psi_1, \psi_n)$, where $n$ refers to the $n$-th neighbor shell in the [100] direction (perpendicular to the rotational axis of $\psi$) and $\psi$ the tilt-angle around the $z$-axis (\autoref{fig:corr}a).
Note that the same correlation occurs for the symmetrically equivalent [010] direction.
This is analysis is carried out for \ce{CsPbI3}, but the behavior is qualitatively the same for all materials considered here (\autoref{fig:all-correlations} in the \gls{sm}).
There is a clear correlation between $\psi_1$ and $\psi_n$ for small $n$, which can be quantified with the Pearson correlation $p_n$ defined as
\begin{align*}
    p_n = \frac{\left < \psi_1 \psi_n \right > }{\left < \psi_1 \psi_1 \right >}.
\end{align*}

The correlation naturally occurs due to the nearest neighbor octahedra sharing one halogen ion ($X$=I, Br, Cl) atom (\autoref{fig:functionals}a).
If a given octahedron rotates in one direction its neighbors will thus favorably rotate in the opposite direction.
This leads to the sign of the correlation alternating between odd and even neighbor shells along the [100] direction.

The absolute value of the correlation $|p_n|$ decays exponentially with a typical length-scale of about 3 to 5 shells (corresponding to about \SI{20}{\angstrom}) with the number of the neighbor shell (\autoref{fig:corr}b).
We note that at \SI{500}{\kelvin} (and thus above the tetragonal-to-cubic transition temperature), there is a  significant correlation between octahedra lying multiple shells apart, stretching as far as the seventh neighbor shell.
Although the correlation weakens at temperatures between 600 and \SI{700} Kelvin, which is far into the cubic stability region, the orientation of a specific octahedron still has a notable effect on its second and third neighbors.

The full temperature dependence of the correlation $p_n$ (\autoref{fig:corr}c) shows that the correlation increases as the temperature decreases towards the transition to the tetragonal phase.
While the tilt angle around the $z$-axis ($\psi$) correlates strongly between neighboring octahedra along the [100] direction, there is almost no correlation in the cubic phase between neighboring octahedra along the [001] direction.\footnote{
    For neighbor shells along the [001] direction the correlation is always positive due to the in-phase tilting of the $a^0a^0c^+$ phase.
}
This is  likely due to the fact that the shared $X$ cation for these neighbors does not move when rotating around the $z$-axis, and thus the octahedra can rotate around the $z$-axis more independently of each other.
In the tetragonal phase the correlation between all neighboring shells approaches the respective limiting values of $\pm1$, indicating that long-ranged ordering is obtained.

The strong short-range ordering observed in the cubic phase may have implication for refinement of experimental diffraction data.
Since some of the tetragonal ordering is still present in the cubic phase, the structure determination can be difficult in these materials.
Specifically, our findings put into question the frequently adopted approach of modelling thermal displacement parameters through Debye-Waller factors in this class of materials.
The latter (as commonly used) incorporate namely the assumption that the local energy landscape can be approximated as harmonic.
Halide perovskites, however, exhibit extreme anharmonicity \cite{carignano2017critical, marronnier2018anharmonicity}, in particular for the zone boundary modes that are connected to the octahedral tilting \cite{FraRosEri22}.

\section{Conclusions}
In conclusion, in this study we have explored the local structural properties of inorganic halide perovskites through large-scale \gls{md} simulations using machine-learned potentials based on \gls{dft} calculations.
By assessing the phase transitions and local tilt angles of octahedra in the perovskite structure, we first analyzed the performance of different functionals as described by \gls{nep} models, including PBE, PBEsol, \gls{scan}, and \gls{cx}.
We find that all models (functionals) underestimate the phase transition temperatures, with the \gls{cx}-based model yielding the values that are closest to experiment.
Focusing subsequently on the \gls{cx} method, we then established trends across a family of \ce{Cs$MX$3} perovskites.
Finally, we demonstrated the presence of strong short-range ordering, reminiscent of the tetragonal phase, even in the cubic phase of halide perovskites for above the cubic-to-tetragonal transition temperature.
This ordering connects the transiently strongly distorted local structure to the global cubic arrangement and sheds light on the complex interplay between local and global structural properties in these materials.
This study provides direct insight into the local disorder in inorganic halide perovskites across a wide temperature range and contributes to the more complete understanding of their structural properties.

\section{Acknowledgments}

This work was funded by the Swedish Research Council (grant numbers 2018-06482, 2019-03993, 2020-04935, 2021-05072), the Area of Advance Nano at Chalmers, and the Chalmers Initiative for Advancement of Neutron and Synchrotron Techniques.
J.~W. acknowledges the Swedish Strategic Research Foundation through a Future Research Leader programme (FFL21-0129).
The computations were enabled by resources provided by the National Academic Infrastructure for Supercomputing in Sweden (NAISS) and the Swedish National Infrastructure for Computing (SNIC) at C3SE, NSC, HPC2N, and PDC partially funded by the Swedish Research Council through grant agreements no. 2022-06725 and no. 2018-05973.

\end{document}